\documentclass[a4paper,12pt]{article}

\usepackage[dvipdfmx]{graphicx}
\usepackage{amsmath}
\usepackage{amssymb}
\usepackage{ulem}
\usepackage{setspace}
\usepackage{cancel}
\usepackage{xcolor} 

\begin{document}
\begin{titlepage}
\null
\begin{flushright}
August, 2016
\end{flushright}

\vskip 1.8cm
\begin{center}

  {\Large \bf Non-geometric Five-branes in Heterotic Supergravity}

\vskip 1.8cm
\normalsize

  {\bf Shin Sasaki\footnote{shin-s(at)kitasato-u.ac.jp} 
and Masaya Yata\footnote{phymasa(at)nus.edu.sg}}

\vskip 0.5cm

  { \it
  Department of Physics \\
  Kitasato University \\
  Sagamihara 252-0373, Japan \\
\vspace{0.5cm}
  Department of Physics, \\
  National University of Singapore \\
  2, Science Drive 3, Singapore 117542, Singapore
  }

\vskip 2cm

\begin{abstract}
We study T-duality chains of five-branes in heterotic supergravity where
 the first order $\alpha'$-corrections are present.
By performing the $\alpha'$-corrected T-duality transformations of the
heterotic NS5-brane solutions, we obtain the KK5-brane and the exotic
 $5^2_2$-brane solutions associated with the symmetric, the neutral and
 the gauge NS5-branes.
We find that the Yang-Mills gauge field in these solutions satisfies the
 self-duality condition in the three- and two-dimensional transverse spaces to the
 brane world-volumes.
The $O(2,2)$ monodromy structures of the $5^2_2$-brane solutions are
 investigated by the $\alpha'$-corrected generalized metric.
Our analysis shows that the symmetric $5^2_2$-brane solution, which satisfies the standard
 embedding condition, is a T-fold and it exhibits the non-geometric nature.
We also find that the neutral $5^2_2$-brane solution is a T-fold at
 least at $\mathcal{O} (\alpha')$.
On the other hand, the gauge $5^2_2$-brane solution is not a T-fold but show unusual structures of space-time.
\end{abstract}
\end{center}
\end{titlepage}

\newpage
\tableofcontents
\setcounter{footnote}{0}

\section{Introduction}
Extended objects such as branes play an important role in superstring
theories.
The U-duality \cite{Hull:1994ys}, which makes non-trivial connections among consistent superstring
theories, relates various branes in each theory.
It was shown that M-theory compactified on $T^d$ has the U-duality
symmetry group $E_{d(d)}(\mathbb{R})$ in lower dimensions 
\cite{Elitzur:1997zn,Obers:1998fb, Blau:1997du}.
BPS branes in superstring theories form lower-dimensional multiplets
under the U-duality group.
For example, when we consider M-theory compactified on $T^8$, we have
the $E_{8(8)} (\mathbb{R})$ BPS point particle multiplet.
The higher dimensional origin of parts of these point particles is the ordinary branes
wrapped/unwrapped on cycles in $T^8$.
Here, the ordinary branes are waves, F-strings, D-branes, NS5-branes,
Kaluza-Klein (KK) branes.
However, there are states whose higher dimensional origin does not trace
back to the ordinary branes.
These states are called exotic states and their higher dimensional origin
is known as exotic branes \cite{Eyras:1999at, LozanoTellechea:2000mc}.

Among other things, an exotic brane in type II string theories
known as the $5^2_2$-brane, has been studied intensively  
\cite{Kimura:2013fda,Kimura:2015yla,Andriot:2014uda,Kimura:2014bea}.
The most tractable duality in string theory is the T-duality.
Type II string theories compactified on $T^d$ has T-duality symmetry group $O(d,d, \mathbb{Z})$.
The exotic $5^2_2$-brane is obtained by performing the T-duality transformations along the
transverse directions to the NS5-brane world-volume.
As its notation suggests, the $5^2_2$-brane has two isometries in the transverse
directions to the brane world-volume \cite{Obers:1998fb}\footnote{In
this paper, we use the notation $5^2_2$ in the sense that they 
are obtained by the T-duality transformations twice on the heterotic NS5-branes.}. 
Its tension is proportional to $g_s^{-2}$ where $g_s$ is the string coupling constant. 
Therefore the $5^2_2$-brane is a solitonic object of co-dimension two.
These co-dimension two objects in string theory, sometimes called defect
branes \cite{Bergshoeff:2011se}, exhibit specific properties \cite{Park:2015gka,Okada:2014wma}.
When one goes around the center of the co-dimension two branes in the transverse
two directions and come back to the original point, the background
geometry of the branes changes according to the non-trivial
monodromy. Therefore the metric and other background fields 
are generically governed by multi-valued functions.
In this sense, they are called non-geometric \cite{Hull:2004in}.
However when the monodromy is given by the symmetry group of the theory, which is
generically the U-duality group in string theory, then the non-geometry becomes healthy candidate of 
solutions to string theory. This kind of solution is called U-fold.
An important nature of the exotic branes is that they are 
non-geometric objects \cite{deBoer:2010ud}.
Therefore they are called non-geometric branes or Q-branes \cite{Hassler:2013wsa}.
In particular, the $5^2_2$-brane in type II string theories is a T-fold, whose monodromy is given by
the T-duality group $O(2,2)$.

The purpose of this paper is to study exotic branes in heterotic string theories.
Compared with type II string theories, exotic branes in type I and
heterotic string theories have been poorly understood.
This is due to the non-Abelian gauge field living in the space-time in these theories.
Notably, the Yang-Mills gauge field enters into the space-time action as the
first order $\alpha'$-corrections. 
Due to the $\alpha'$-corrections, the Buscher rule \cite{Buscher:1987sk}
of the T-duality transformation for heterotic supergravity 
is modified \cite{Tseytlin:1991wr, Bergshoeff:1995cg, Serone:2005ge}.
The most famous extended objects in heterotic supergravity theories are
the heterotic NS5-brane solutions.
There are three distinct NS5-brane solutions in heterotic supergravity 
\cite{Strominger:1990et,Callan:1991dj,Duff:1990wv}.
We will perform the T-duality transformations to these heterotic
NS5-branes by the $\alpha'$-corrected Buscher rule and obtain new
five-brane solutions.
We will then study the monodromy structures of these solutions.

The organization of this paper is as follows.
In the next section, we introduce the heterotic NS5-brane solutions 
known as the symmetric, neutral and gauge types 
\cite{Strominger:1990et,Callan:1991dj}.
In section 3, we introduce the isometries along the transverse
directions to the NS5-brane world-volumes and 
perform the T-duality transformation by the $\alpha'$-corrected
Buscher rule.
We obtain the heterotic Kaluza-Klein five-brane (KK5-brane) solutions associated with the
three types of solutions.
We then perform the second T-duality transformations on the KK5-branes 
and write down the exotic $5^2_2$-brane solutions.
In section 4, we examine the monodromy structure of the heterotic
$5^2_2$-brane solutions by the $\alpha'$-corrected generalized metric.
We show that the monodromy is given by the $O(2,2)$
T-duality group for the symmetric and the neutral solutions while the
gauge solution remains geometric.
Section 5 is devoted to conclusion and discussions.
The explicit form of the smeared gauge KK5-brane solution is found in
appendix.

\section{Heterotic NS5-brane solutions}
In this section, we introduce the NS5-brane solutions in ten-dimensional
heterotic supergravity which is the low-energy effective theory of heterotic
superstring theory.  
Heterotic supergravity consists of the ten-dimensional $\mathcal{N}
= 1$ gravity multiplet coupled with the $\mathcal{N} = 1$ vector multiplet.
The relevant bosonic fields in heterotic supergravity are the vielbein
$e_M {}^A$, the dilaton $\phi$, the NS-NS $B$-field $B_{MN}$ and the
Yang-Mills gauge field $A_M$.
Here $M,N, = 0, \ldots, 9$ are the curved space indices while $A, B, \ldots = 0, 1,
\ldots, 9$ are the local Lorentz indices.
The Yang-Mills gauge field $A_M$ is in the adjoint representation of 
the gauge group $G$ which is $SO(32)$ or $E_8 \times E_8$.
We employ the convention such that the gauge field is represented by
anti-hermitian matrices and the trace is taken over the matrices of the
fundamental representation. 

A remarkable property of heterotic supergravity is that the Yang-Mills gauge
field contributes to the action as the first order $\alpha'$-correction.
The famous anomaly cancellation mechanism and the supersymmetry
completion result in the Riemann curvature square
term which involves higher derivative corrections in the same order in $\alpha'$
\cite{Bergshoeff:1988nn}\footnote{The authors
would like to thank Tetsuji Kimura for his introducing \cite{Bergshoeff:1988nn}.}.
The ten-dimensional heterotic supergravity action for the bosonic
fields at the first order in $\alpha'$ is given by
\begin{align}
S =& \ \frac{1}{2 \kappa^2_{10}} \int \! d^{10} x \sqrt{-g} e^{-2\phi} 
\Bigl[
R (\omega) - \frac{1}{3} \hat{H}_{MNP}^{(3)} \hat{H}^{(3)MNP} + 4 \partial_M \phi \partial^M
 \phi 
\notag \\
& 
\qquad 
\qquad 
\qquad 
\qquad 
\qquad 
 \frac{}{}
+ \alpha' 
\left(
\text{Tr} F_{MN} F^{MN} + R_{MN AB} (\omega_{+}) R^{MNAB} (\omega_{+})
\right)
\Bigr].
\label{eq:heterotic_sugra}
\end{align}
Here we employ the convention such that $\frac{\kappa^2_{10}}{2g^2_{10}} =
\alpha'$ where $\kappa_{10}$ and $g_{10}$ are the gravitational
and the gauge coupling constants in ten dimensions.
The ten-dimensional metric $g_{MN}$ in the string frame is defined
through the vielbein as $g_{MN} = \eta_{AB} e_M {}^A e_N {}^B$
where the metric in the local Lorentz frame is $\eta_{AB} = \text{diag} (-1, 1, \ldots, 1)$.
The Ricci scalar $R (\omega)$ is constructed from
the spin connection $\omega_{M} {}^{AB}$:
\begin{align}
R (\omega) =& \ e^M {}_A e^N {}_B R^{AB} {}_{MN} (\omega), \notag \\
R^{AB} {}_{MN} (\omega) =& \ \partial_M \omega_N {}^{AB} - \partial_N
 \omega_M {}^{AB} + \omega_M {}^{AD} \eta_{CD} \omega_N {}^{CB} -
 \omega_N {}^{AD} \eta_{CD} \omega_M {}^{CB}.
\end{align}
The spin connection is expressed by the vielbein $e_M {}^A$ and its
inverse:
\begin{align}
\omega_{M}^{~AB}={1\over2} 
\Bigl[
e^{AN}( \partial_{M}e_{N}^{~B} - \partial_{N}e_{M}^{~B} )
-e^{BN}( \partial_{M}e_{N}^{~A} - \partial_{N}e_{M}^{~A} )
-e^{AP} e^{BQ}
( \partial_{P} e_{Q C} - \partial_{Q}e_{P C} ) 
e_{M}^{~C}
\Bigr].
\end{align}
The modified spin connection $\omega_{\pm M} {}^{AB}$ 
which enters into the action in the $\mathcal{O} (\alpha')$ terms 
is defined as 
\begin{align}
\omega_{\pm M} {}^{AB} = \omega_{M} {}^{AB} \pm \hat{H}_{M}^{(3)} {}^{AB},
\label{eq:modified_spin_connection}
\end{align}
where 
$\hat{H}_M^{(3)} {}^{AB} = e^{NA} e^{PB} \hat{H}_{MNP}^{(3)}$
 and the modified $H$-flux $\hat{H}_{MNP}^{(3)}$ is defined by 
\begin{align}
\hat{H}^{(3)}_{MNP} =& \ H_{MNP}^{(3)} + \alpha' 
\left(
\Omega_{MNP}^{\text{YM}}
- 
\Omega_{MNP}^{\text{L}+}
\right).
\label{eq:H-flux}
\end{align}
Here the field strength $H_{MNP}^{(3)}$ of the $B$-field is defined by 
\begin{align}
H_{MNP}^{(3)} =& \ \frac{1}{2} (\partial_M B_{NP} + \partial_N B_{PM} +
 \partial_P B_{MN}).
\end{align}
The Yang-Mills and the Lorentz Chern-Simons terms which appear in \eqref{eq:H-flux}
are defined as follows:
\begin{align}
\Omega^{\text{YM}}_{MNP} =& \ 
3! \text{Tr}
\left(
A_{[M} \partial_N A_{P]} + \frac{2}{3} A_{[M} A_{N} A_{P]}
\right), \notag \\
\Omega^{\text{L}+}_{MNP} =& \ 
3!
\left(
\eta_{BC} \eta_{AD}\omega_{+[M} {}^{AB} \partial_{N} \omega_{+P]} {}^{CD} 
+ \frac{2}{3} \eta_{AG} \eta_{BC} \eta_{DF} \omega_{+[M}^{AB} \omega_{+N} {}^{CD}
 \omega_{+P]} {}^{FG}
\right).
\end{align}
Here the symbol $[M_1 M_2 \cdots M_n]$ stands for the anti-symmetrization of
indices with weight $\frac{1}{n!}$. 
Note that the modified $H$-flux is iteratively defined through the
relations \eqref{eq:modified_spin_connection} and \eqref{eq:H-flux}
order by order in $\alpha'$. 
The modified $H$-flux $\hat{H}^{(3)}$ obeys the following Bianchi identity:
\begin{align}
d \hat{H}^{(3)} = \alpha'
\left(
\text{Tr} F \wedge F - \text{Tr} R \wedge R
\right) + \mathcal{O} (\alpha^{\prime 2}),
\label{eq:Bianchi}
\end{align}
where $R^{AB} = \frac{1}{2!} R^{AB} {}_{MN} d x^M \wedge d x^N$ is the
$SO(1,9)$-valued curvature 2-form.
The component of the Yang-Mills gauge field strength 2-form 
$F = \frac{1}{2!} F_{MN} d x^M \wedge dx^N$
 is 
\begin{align}
F_{MN} {}^I = \partial_M A_N^I - \partial_N A_M^I + f^I {}_{JK} A^J
 {}_M A^K {}_N.
\end{align}
Here $I,J,K = 1, \ldots, \text{dim} \mathcal{G}$ are the gauge indices 
and $f^I {}_{JK}$ is the structure constant for the Lie algebra $\mathcal{G}$ associated with $G$.
Note that the terms $R_{ABMN} (\omega_+) R^{ABMN} (\omega_+)$ and
$\Omega^{\text{L}+}_{MNP}$ in the action \eqref{eq:heterotic_sugra} are
higher derivative corrections to the ordinary second order derivative terms.

The heterotic NS5-brane solutions satisfy the equation of motion
 derived from the action \eqref{eq:heterotic_sugra} at $\mathcal{O}
 (\alpha')$.
There are three distinct NS5-brane solutions known as the symmetric, the
neutral and the gauge types in heterotic supergravity
\cite{Strominger:1990et,Callan:1991dj}. 
They are 1/2 BPS configurations and preserve a half of the sixteen
 supercharges.
They satisfy the following ansatz:
\begin{align}
& \ ds^2 = \eta_{ij} dx^i dx^j + H (r) \delta_{mn} dx^m dx^n, \notag \\
& \ \hat{H}^{(3)}_{mnp} = - \frac{1}{2} \varepsilon_{mnpq} \partial_q H (r), \qquad 
\phi = \frac{1}{2} \log H (r), \notag \\
& \ r^2 = (x^1)^2 + (x^2)^2 + (x^3)^2 + (x^4)^2,
\label{eq:five-branes}
\end{align}
where the indices $i,j = 0,5,6,7,8,9$, $m,n = 1,2,3,4$ represent the
world-volume and the transverse directions to the NS5-branes
and $\varepsilon_{mnpq}$ is the Levi-Civita symbol.
The 1/2 BPS condition leads to the self-duality condition for the
Yang-Mills gauge field:
\begin{align}
F_{mn} = \tilde{F}_{mn},
\label{eq:self-dual}
\end{align}
and other components $A_i$ vanish. 
Here the Hodge dual field strength in the transverse four-dimensions is
defined by $\tilde{F}_{mn} = \frac{1}{2} \varepsilon_{mnpq} F^{pq}$.
The harmonic function $H(r)$ is determined by the Yang-Mills gauge field
configurations of the solutions.
These heterotic NS5-branes are characterized by two charges.
One is the topological charge $k$ associated with the gauge instantons:
\begin{align}
k = - \frac{1}{ 32 \pi^2} \int \mathrm{Tr} [F \wedge F],
\end{align}
where the integral is defined in the transverse four-space.
The other is the charge $Q$ associated with the 
modified $H$-flux:
\begin{align}
Q = - \frac{1}{2\pi^2 \alpha'} \int_{S^3} \! \hat{H}^{(3)}.
\end{align}
Here $S^3$ is the asymptotic three-sphere surrounding the NS5-brane.
In the following we briefly introduce the three distinct NS5-brane solutions.

\paragraph{Symmetric solution}
The most tractable NS5-brane solution in heterotic theory may be 
the so-called symmetric solution. 
The Yang-Mills gauge field which satisfies the self-duality condition \eqref{eq:self-dual} is given by an
instanton solution in four-dimensions.
The gauge field takes value in the $SU(2)$ subgroup of the gauge
group $G$ and the explicit solution is given by \cite{Callan:1991dj},
\begin{align}
A_m =& \ - \frac{ \sigma_{mn} x^n}{r^2 + n \alpha' e^{-2\phi_0}}, 
\quad
n \in \mathbb{Z}, 
\notag \\
H (r) =& \ e^{2\phi_0} + \frac{n \alpha'}{r^2},
\label{eq:symmetric}
\end{align}
where $\sigma_{mn}$ is the self-dual part of the $SO(4)$ Lorentz
generator.
This is written in the following form,
\begin{align}
\sigma_{mn} = \eta^I_{mn} T^I, ~~~ \eta^I_{mn} = 
 \varepsilon^I_{~mn4} + \delta^I_{~m} \delta_{n4} - \delta^I_{~n}
 \delta_{m4}, 
\end{align}
where $\eta^I_{mn}$ is the 't Hooft symbol which satisfies the self-duality condition in terms of the indices
$m,n$:
\begin{align}
 \eta^I_{mn} = {1\over2} \varepsilon_{mnpq}
 \eta^I_{pq}. 
\end{align}
The anti-hermitian matrices $T^I$ are defined as 
\begin{align}
T^1 = 
\left[
\begin{array}{cccc}
0 & 0 & 0 & -1 \\
0 & 0 & 1 & 0 \\
0 & -1 & 0 & 0 \\
1 & 0 & 0 & 0
\end{array}
\right], \quad 
T^2 = 
\left[
\begin{array}{cccc}
0 & 0 & -1 & 0 \\
0 & 0 & 0 & -1 \\
1 & 0 & 0 & 0 \\
0 & 1 & 0 & 0
\end{array}
\right], \quad 
T^3 = 
\left[
\begin{array}{cccc}
0 & 1 & 0 & 0 \\
-1 & 0 & 0 & 0 \\
0 & 0 & 0 & -1 \\
0 & 0 & 1 & 0
\end{array}
\right].
\end{align}
They satisfy the $su(2)$ algebra
\begin{align}
[T^I, T^J] = - 2 
\varepsilon^{IJK}
 T^K, \quad (I,J,K = 1,2,3).
\end{align}
The solution (\ref{eq:symmetric}) is nothing but
the BPST one-instanton in the non-singular gauge \cite{Belavin:1975fg}.
The constant $\phi_0$ is the asymptotic value of the dilaton.
Compared with the general BPST instanton solution, the symmetric
solution has a fixed finite instanton size $\rho = e^{-\phi_0} \sqrt{n \alpha'}$.
This solution has charges $(k,Q) = (1, n)$.
A remarkable fact about the symmetric solution 
is that the dilaton configuration, hence the harmonic function $H(r)$, 
is obtained through the standard embedding ansatz:
\begin{align}
(A_m)^{ab} = \omega_{+m} {}^{ab}.
\label{eq:standard_embedding}
\end{align}
Here the $SU(2) \subset G$ indices $a,b$ of the gauge field are
identified with the indices of the $SU(2)$ subgroup of the local Lorentz
group $SO(4)$ in the transverse directions.
If the relation \eqref{eq:standard_embedding} holds, the Chern-Simons
terms in the modified $H$-flux \eqref{eq:H-flux} cancel out.
Therefore the $H$-flux only comes from the $B$-field and the 
Bianchi identity \eqref{eq:Bianchi} becomes $d \hat{H}^{(3)} = 0$.
From the Bianchi identity and the relation between $H(r)$ and
$\hat{H}_{mnp}^{(3)}$ in \eqref{eq:five-branes}, 
the $B$-field is determined through the following condition:
\begin{align}
\partial_m H = {1\over2} 
\varepsilon_{mnpq}
  \partial_{n} B_{pq}.
\label{eq:B_symmetric}
\end{align}
Although the symmetric NS5-brane solution \eqref{eq:five-branes} with 
\eqref{eq:symmetric} is a solution to the equation of motion derived from the action
\eqref{eq:heterotic_sugra} at $\mathcal{O} (\alpha')$, the string sigma-model analysis indicates that 
the symmetric solution is an exact solution valid at all orders in $\alpha'$ \cite{Callan:1991dj}.
Indeed, the symmetric solution has the $\mathcal{N} = (4,4)$ sigma model
description and it is protected against $\alpha'$-corrections.

\paragraph{Neutral solution}
The neutral solution is the one with charges $(k,Q) = (0,n)$ \cite{Callan:1991dj}.
For this solution, the Yang-Mills gauge field becomes trivial and the harmonic
function is given by
\begin{align}
A_M =& \ 0, \notag \\
H (r) =& \ e^{2\phi_0} + \frac{n \alpha'}{r^2}.
\label{eq:neutral}
\end{align}
Since the gauge field does not appear in the neutral solution, 
this is a solution to type II supergravities.
Indeed, this is the type II NS5-brane solution.
It is remarkable that the neutral solution \eqref{eq:five-branes} with 
\eqref{eq:neutral} for heterotic supergravity is valid at the first order in $\alpha'$.
For this solution, we find the curvature $R_{MNAB}$ is order
$\mathcal{O}(\alpha')$.
Therefore the Lorentz Chern-Simons term in the modified $H$-flux
\eqref{eq:H-flux} becomes a higher order correction in $\alpha'$ and negligible.
Then the Bianchi identity again becomes $d\hat{H}^{(3)} = 0$.
With the solution \eqref{eq:neutral} at hand, the $B$-field is given by
that of the type II NS5-brane 
and its components, which are determined by \eqref{eq:B_symmetric}, 
are the same as the ones of the symmetric solution in the linear
order in $\alpha'$.
We note that the neutral solution has the $\mathcal{N} = (4,0)$ worldsheet sigma model description
and receives higher order $\alpha'$-corrections in heterotic theories \cite{Callan:1991dj}.

\paragraph{Gauge solution}
The last is the so called gauge solution which has been originally
found in \cite{Strominger:1990et}.
The Yang-Mills gauge field is again given by the BPST instanton.
As in the case of the neutral solution, the curvature term in the
modified $H$-flux is neglected as it is a higher order in $\alpha'$.
However, the Yang-Mills gauge field still contributes to the $H$-flux in the gauge
solution and the Bianchi identity becomes $d \hat{H}^{(3)} = \alpha' \text{Tr} F \wedge F$.
From the Bianchi identity, the harmonic function is determined to be
\begin{align}
A_m =& \ -\frac{ \sigma_{mn} x^n}{r^2 + \rho^2}, 
\notag \\
H (r) =& \ e^{2\phi_0} + 8 \alpha' \frac{r^2 + 2 \rho^2}{(r^2 +
 \rho^2)^2}.
\label{eq:gauge_solution}
\end{align}
Compared with the symmetric solution \eqref{eq:symmetric}, the size modulus $\rho$ is not
fixed for the gauge solution and it has charges $(k,Q) = (1,8)$.
Similar to the neutral solution, the gauge solution 
is obtained by a perturbative series of $\alpha'$ and the functional
forms \eqref{eq:five-branes}, \eqref{eq:gauge_solution} are 
valid at the first order in $\alpha'$. 
Indeed, the gauge solution has the $\mathcal{N} = (4,0)$ worldsheet sigma model description 
and is expected to receive higher order $\alpha'$-corrections
\cite{Callan:1991dj}.
The position of the instanton corresponds to that of the NS5-brane.
Therefore the instanton with finite size $\rho \not= 0$ resides in the
core of the NS5-brane.  
It has been discussed in \cite{Witten:1995gx}, when the instanton shrink
to zero size $\rho \to 0$, a gauge multiplet on the brane world-volume
becomes massless and the $SU(2)$ gauge symmetry is enhanced.
The gauge NS5-brane in the $SO(32)$ heterotic theory
in the zero size limit $\rho \to 0$ is related to
the D5-brane in type I theory by the S-duality.

We make a comment on the $B$-field for the gauge NS5-brane solution.
A careful analysis reveals that only the Yang-Mills Chern-Simons term
contributes to the modified $H$-flux and we find $H^{(3)}_{MNP} = 0$ at least at $\mathcal{O} (\alpha')$. 
Therefore the $B$-field is not excited in the gauge NS5-brane solution.
Then, the components of the $B$-field take a constant value:
\begin{align}
B_{mn} = \Theta_{mn},~~~(\mbox{constant} ).
\end{align}

\section{T-duality chains of five-branes and $\alpha'$-corrected Buscher
 rule}
In this section, we derive new five-brane solutions in heterotic
supergravity in the family of the T-duality chains.
Since the gauge field enters into the action \eqref{eq:heterotic_sugra} as the
first order $\alpha'$-correction, the T-duality transformation in the
heterotic supergravity should be modified from the standard Buscher rule
\cite{Buscher:1987sk}.
The first order $\alpha'$-corrections to the Buscher rule 
in heterotic supergravity 
have been written down in \cite{Bergshoeff:1995cg}. This is given by
\begin{align}
g_{\hat{M} \hat{N}} \ \longrightarrow& \ g_{\hat{M} \hat{N}} + \frac{1}{G'_{yy}} 
\left(
g_{yy} G'_{y\hat{M}} G'_{y\hat{N}} - G'_{yy} g_{y\hat{M}} G'_{y\hat{N}} - G'_{yy} G'_{y\hat{M}} g_{y\hat{N}}
\right), 
\notag \\
g_{y \hat{M}} \ \longrightarrow& \ - \frac{g_{y\hat{M}}}{G'_{yy}} + \frac{g_{yy}
 G'_{y\hat{M}}}{G^{\prime 2}_{yy}}, \qquad 
g_{yy} \ \longrightarrow \ \frac{g_{yy}}{G^{\prime 2}_{yy}}, 
\notag \\
B_{\hat{M}\hat{N}} \ \longrightarrow& \ B_{\hat{M}\hat{N}} +
 \frac{1}{G'_{yy}} (G'_{y\hat{M}} B_{\hat{N}y} - G'_{y\hat{N}}
 B_{\hat{M}y}), \qquad 
B_{y\hat{M}} \ \longrightarrow \ - \frac{B_{y\hat{M}}}{G'_{yy}} - \frac{G'_{y\hat{M}}}{G'_{yy}},
\notag \\
\phi \ \longrightarrow& \ \phi - \frac{1}{2} \log |G'_{yy}|, \qquad 
A_{\hat{M}}^I \ \longrightarrow \ A_{\hat{M}}^I -
 \frac{G'_{y\hat{M}}}{G'_{yy}} A_y^I, \qquad 
A_y^I \ \longrightarrow \ - \frac{A_y^I}{G'_{yy}},
\label{eq:Buscher}
\end{align}
where we have decomposed the indices $M = (\hat{M}, y)$.
The indices $y$ and $\hat{M}, \hat{N} \ldots$ specify an isometry and
non-isometry directions respectively.
In \eqref{eq:Buscher}, we have defined the following quantity,
\begin{align}
G'_{MN} = g_{MN} - B_{MN}  + 2 \alpha' 
\bigl[
\mbox{Tr} ( \omega_{+M} \omega_{+ N} ) -  \mbox{Tr} ( A_M A_N ) 
\bigr],
\label{eq:Buscher_G}
\end{align}
where we have defined $ \mbox{Tr} \omega_{+M} \omega_{+ N} = \omega_{+M} {}^{AB} \omega_{+ NBA} $.
We call \eqref{eq:Buscher} 
with \eqref{eq:Buscher_G} 
the heterotic Buscher rule.
Note that when the Yang-Mills gauge field $A_M {}^I$ and the higher derivative
corrections coming from $\omega_{+M} {}^{AB}$ are turned off, the
relation \eqref{eq:Buscher} reduces to the ordinary Buscher rule for
the NS-NS backgrounds in type II supergravities.

\subsection{Heterotic KK5-branes}
Before going to the exotic $5^2_2$-branes, we first write down the
heterotic KK5-brane solutions.
In order to perform the T-duality transformation for the heterotic NS5-brane
solutions, we introduce a $U(1)$ isometry along the transverse
direction to the brane world-volumes \footnote{If we perform the T-duality
transformations along the world-volume direction of the NS5-branes in
the $SO(32)$ heterotic theory, we obtain the identical solutions in the $E_8
\times E_8$ theory and vice versa.}.
To this end, we first compactify the $x^4$-direction with the radius $R_4$
and consider the periodic array of the NS5-branes.
Then by taking the small radius limit $R_4 \to 0$, we introduce the
$U(1)$ isometry to the heterotic NS5-brane solutions.
The self-duality condition \eqref{eq:self-dual} reduces to the monopole
equation 
$\frac{1}{2} \varepsilon_{m'n'p'} F_{n'p'} =
\partial_{m'} A_4 + [A_{m'}, A_4]$ where $m',n',p' = 1,2,3$. 
Therefore the resulting solutions are
called the heterotic monopoles or smeared NS5-branes.
The explicit forms of the smeared NS5-brane solutions which originate
from the periodic arrays of the symmetric, the neutral and the gauge solutions have been written down in 
\cite{Khuri:1992hk, Gauntlett:1992nn}.
By performing the T-duality transformation of the solutions along the
isometry direction, we obtain the KK5-brane solutions associated
with the three types of the NS5-branes.
In the following, we calculate the T-duality transformations of the
NS5-brane solutions and derive the KK5-brane solutions.

\paragraph{Symmetric KK5-brane}
For the symmetric solution where the standard embedding condition is
satisfied, the ansatz and the Bianchi identity leads to the condition
$\Box e^{2\phi} = \Box H = 0$.
The periodic array of the symmetric NS5-brane solution is governed by
the following harmonic function $H (r)$ on $\mathbb{R}^3 \times S^1$
\cite{Gauntlett:1992nn}:
\begin{align}
H (r) = e^{2\phi_0} + \sum_{s} \frac{n \alpha'}{r^2 + (x^4 - 2\pi R_4 s)^2}, \quad 
r^2 = (x^1)^2 + (x^2)^2 + (x^3)^2.
\label{eq:harmonic_sum}
\end{align}
Taking the compactification radius small $R_4 \to 0$, 
the sum in \eqref{eq:harmonic_sum} is approximated by the integral
over $s$. We call this procedure as smearing.
The result is\footnote{
The constant $C = n \alpha' R_4^{-1}$ would diverge in the limit $R_4
\to 0$ but this is an artifact of the smearing procedure. We can
find the harmonic function \eqref{eq:monopole_H} which finite $C$ by solving the Laplace
equation $\Box H = 0$ in three dimensions.
}
\begin{align}
H (r) = e^{2\phi_0} + \frac{C}{2r},
\label{eq:monopole_H}
\end{align}
The corresponding solution is the smeared symmetric NS5-brane of co-dimension
three discussed in \cite{Khuri:1992hk}.
We note that the smeared symmetric NS5-brane is an H-monopole whose
quantized charge $Q$ is well-defined.
On the other hand, the Yang-Mills monopole charge for the smeared solution is
not defined anymore \cite{Gauntlett:1992nn}.
We now perform the T-duality transformation on the smeared symmetric NS5-brane solution.
For a solution where the standard embedding 
\eqref{eq:standard_embedding} is satisfied, the heterotic 
Buscher rule \eqref{eq:Buscher} is quite simplified.
This is because the relation \mbox{Tr}( $\omega_{+M}  \omega_{+ N} )= \mbox{Tr} (A_M A_N)$
holds and the last term in \eqref{eq:Buscher_G} vanishes.

We perform the T-duality transformation by utilizing the heterotic Buscher rule \eqref{eq:Buscher}.
Then we obtain the symmetric KK5-brane solution:
\begin{align}
ds^2 =& \ \eta_{ij} dx^i dx^j + H 
\left[
(dx^1)^2 + (dx^2)^2 + (dx^3)^2
\right] 
+ H^{-1}
\left[
dx^4 + \beta_{m'} dx^{m'}
\right]^2, \notag \\
\phi =& \ 0, \qquad B_{MN} = 0, \qquad 
(m'=1,2,3), 
\label{eq:KKM_metric}
\end{align}
where the harmonic function $H$ is given in \eqref{eq:monopole_H}
 and $\beta_{m'}$ is determined by the following relation:
\begin{align}
\partial_{m'} H = 
\varepsilon_{m'n'p'}
 \partial_{n'} \beta_{p'}. 
\label{betaRelation}
\end{align}
The dilaton and the NS-NS $B$-field vanish but the Yang-Mills gauge field
remains non-trivial.
The symmetric KK5-brane solution is not a purely geometric solution but the geometry is dressed up with the Yang-Mills gauge fields.
The gauge field $A_M {}^I$ is obtained as 
\begin{align}
A_1 &= {1\over2 H^2} \bigl[  -\beta_1 \partial_1H T^1 
                                                        -(H \partial_3 H + \beta_1 \partial_2 H )T^2
                                                        +(H \partial_2 H - \beta_1 \partial_3 H)T^3  \bigr], \nonumber\\
A_2 &= {1\over2 H^2} \bigl[   ( H  \partial_3H  -\beta_2\partial_1 H )T^1 
                                                        -\beta_2 \partial_2 H  T^2
                                                        -( H \partial_1 H + \beta_2 \partial_3 H )T^3 \bigr],  \nonumber\\
A_3 &= {1\over2 H^2} \bigl[  -( H \partial_2 H + \beta_3 \partial_1 H )T^1
                                                        +( H \partial_1 H  - \beta_3 \partial_2 H )T^2
                                                        -\beta_3 \partial_3 H  T^3\bigr], \nonumber\\
A_4 &= {1\over2 H^2} \bigl[ -\partial_1H T^1 -\partial_2 H T^2 -\partial_3 H T^3\bigr], ~~~~~A_i = 0 .
\label{eq:symmetric_KKM_gauge}
\end{align}
We find that the Yang-Mills gauge field configuration
\eqref{eq:symmetric_KKM_gauge} satisfies the 
anti-monopole (Bogomol'nyi) equation 
\begin{align}
B_i = - D_i \phi.
\label{eq:anti-monopole}
\end{align}
This equation is nothing but the anti-self-duality condition $F_{mn} = -
\tilde{F}^{mn}$ in disguise\footnote{
The flip of the sign in the Hodge dualized field strength
$\tilde{F}_{mn}$ comes from the choice of the freedom for the overall
sign in the heterotic Buscher rule \eqref{eq:Buscher}.
If we choose another 
sign in front of $A_y^I$ in the right hand side of 
\eqref{eq:Buscher}, the gauge field satisfies the self-duality condition
instead of the anti-self-duality condition.
}.
Since the field configuration \eqref{eq:symmetric_KKM_gauge} satisfies
the equation $F_{mn} = - \tilde{F}_{mn}$, this belongs to the general
class of spherically symmetric solutions discussed in \cite{Protogenov:1977tq}.

Meanwhile, by using the relation (\ref{betaRelation}), 
the modified spin connection $\omega_{+}$ associated with the geometry
\eqref{eq:KKM_metric} is calculated to be
\begin{align}
\omega_{+1} &= {1\over2 H^2} \bigl[  -\beta_1 \partial_1H T^1 
                                                        -(H \partial_3 H + \beta_1 \partial_2 H )T^2
                                                        +(H \partial_2 H - \beta_1 \partial_3 H)T^3  \bigr], \nonumber\\
\omega_{+2} &= {1\over2 H^2} \bigl[   ( H  \partial_3H  -\beta_2\partial_1 H )T^1 
                                                        -\beta_2 \partial_2 H  T^2
                                                        -( H \partial_1 H + \beta_2 \partial_3 H )T^3 \bigr],  \nonumber\\
\omega_{+3} &= {1\over2 H^2} \bigl[  -( H \partial_2 H + \beta_3 \partial_1 H )T^1
                                                        +( H \partial_1 H  - \beta_3 \partial_2 H )T^2
                                                        -\beta_3 \partial_3 H  T^3\bigr], \nonumber\\
\omega_{+4} &= {1\over2 H^2} \bigl[ -\partial_1H T^1 -\partial_2 H T^2 -\partial_3 H T^3\bigr], ~~~~~\omega_{+i} = 0. 
\label{eq:symmetric_KKM_omega}
\end{align}
From the results \eqref{eq:symmetric_KKM_gauge} and
\eqref{eq:symmetric_KKM_omega}, we find that the standard embedding
condition \eqref{eq:standard_embedding} still holds for the symmetric KK5-brane solution.

\paragraph{Neutral KK5-brane}
We next study the neutral KK5-brane solution.
For the neutral NS5-brane solution, we have the trivial Yang-Mills gauge field
$A_M {}^I = 0$. 
Again, the smearing procedure is applicable to the neutral solution.
The smeared neutral NS5-brane solution is governed by the harmonic function \eqref{eq:monopole_H}.
As we have claimed in the previous section, the modified spin connection $\omega_+$ is
in the $\mathcal{O}(\alpha')$ for the neutral solution.
Therefore it is negligible in $G'_{MN}$ in the heterotic Buscher rule \eqref{eq:Buscher_G}.
Applying the heterotic Buscher rule, we find that the metric, the $B$-field
and the dilaton for the neutral KK5-brane solution are given by
\eqref{eq:KKM_metric} and the gauge field remains vanishing $A_M
{}^I = 0$. Therefore the neutral KK5-brane is a purely
geometric Taub-NUT solution at least at the leading order in $\alpha'$. 
Indeed, this is nothing but the KK5-brane solution in type II theories.

\paragraph{Gauge KK5-brane}
For the gauge NS5-brane solution, the periodic harmonic function
\eqref{eq:monopole_H} does not work as a solution since $\Box e^{2\phi}
= 0$ is not satisfied for the gauge solution.
The ansatz \eqref{eq:five-branes} and the Bianchi identity indicates 
$\Box H (r) = \frac{\alpha'}{2} \mathrm{Tr} [\varepsilon^{mnpq} F_{mn} F_{pq}]$. 
In order to find a solution of co-dimension three 
associated with the gauge solution, 
we perform the singular gauge transformation of the solution \eqref{eq:gauge_solution}.
Then the gauge field becomes
\begin{align}
A_m = \frac{-2 \rho^2 \bar{\sigma}_{mn} x^n}{r^2 (r^2 + \rho^2)}, \qquad 
r^2 = (x^1)^2 + (x^2)^2+ (x^3)^2 + (x^4)^2, 
\label{eq:gauge_singular}
\end{align}
where $\bar{\sigma}_{mn}$ is the antisymmetric and anti-self-dual matrix and is written as
\begin{align}
\bar{\sigma}_{mn} = \overline{\eta}^I_{mn} T^I,~~~
\bar{\eta}^I_{mn} = 
\varepsilon^I_{~mn4}
 - \delta^I_{~m} \delta_{n 4} + \delta^I_{~n} \delta_{m 4}.  
\end{align}
This is the BPST instanton in the singular gauge. The solution is
obtained by the famous 't Hooft ansatz:
\begin{align}
A_m = \bar{\sigma}_{mn} \partial_n \log f.
\label{eq:tHooft}
\end{align}
The function $f$ 
satisfies 
$\frac{1}{f} \Box f = 0$
and given by 
$f (r) = 1 +  \sum_{i=1}^k \frac{\rho^2}{(x -
x_i)^2}$ for the $k$-instanton solution. 
Here $x_i$ are the positions of the instantons.
Note that the solution \eqref{eq:gauge_singular} corresponds to $k=1$.
Using this fact, we can perform the smearing procedure for $f$ along the line of
obtaining \eqref{eq:monopole_H}.
This periodic array of the instantons is just the calorons of 
the Harrington-Shepard type \cite{Harrington:1978ve}.
The harmonic function $H$ for the smeared gauge NS5-brane is determined
by the Laplace equation whose source is given by the $R_4 \to 0$ limit
of the calorons, 
namely, the smeared instantons. 
After the smearing, we find $f = 1 + \frac{\tilde{\rho}}{r}$ where
$r^2 = (x^1)^2 + (x^2)^2+ (x^3)^2$ and 
$\tilde{\rho} = \frac{\rho^2}{2R_4}$ is a rescaled size modulus.
Then the smeared gauge NS5-brane solution is found to be
\begin{align}
ds^2 &= \eta_{ij} dx^i dx^j + e^{2\phi} \delta_{mn} dx^m dx^n, \nonumber\\
A_{m} &= \bar{\sigma}_{mn} {  -  \tilde{\rho}
 x^{\nu}  \over r^2 (r +  \tilde{\rho}) }, ~~~~~
e^{2\phi} = e^{2\phi_0} -2 \alpha' {   \tilde{\rho}^2 \over r^2 ( r +
 \tilde{\rho})^2 } ,~~~~~
\hat{H}_{mnp}^{(3)} = -{1\over2}   \varepsilon_{mnpq}   \partial_q e^{2\phi}. \label{SmearedD}
\end{align}
As in the case of the gauge NS5-brane solution, we find that the
$B$-fields in the smeared gauge NS5-brane solution becomes trivial.
This property is also found in the heterotic monopole solution
\cite{Gauntlett:1992nn}.
We also note that the smeared gauge NS5-brane 
does not exhibit the H-monopole property.
Indeed, we find the modified $H$-flux behaves like 
$\hat{H}^{(3)}_{mn4} \sim 4 \alpha' \tilde{\rho} \epsilon_{mnq4}
\frac{x^q}{r^6} \ (r \to \infty)$ and the charge $Q$ vanishes.
This is in contrast to the symmetric and the neutral solutions. 
We also comment that the solution \eqref{SmearedD} based on the smeared
instanton does not have a finite monopole charge.
This is again in contrast to the monopole solution in \cite{Gauntlett:1992nn}.

Since the $B$-field is trivial 
in the gauge solution, we can set these components as  
\begin{align}
B_{14}=\Theta_1,~~~
B_{24}=\Theta_2,~~~
B_{34}=\Theta_3,
\end{align}
where $\Theta_i$ are constants. 
The other components do not appear in the solution and they can be set to zero.
Now we perform the T-duality transformation of the smeared gauge
NS5-brane solution along the $x^4$-direction. 
After calculations, we find the following gauge KK5-brane solution:
\begin{align}
ds^2 &= \eta_{ij} dx^i dx^j + {\cal H} \delta_{m'n'} dx^{m'} dx^{n'} +
 e^{-4\phi_0} {\cal H} ( dx^4 + \Theta_{m'} dx^{m'} )^2, \nonumber \\
\phi &= {1\over2} \log e^{-4\phi_0} {\cal H},~~~~~{\cal H}= e^{2\phi_0} -2 \alpha' {\tilde{\rho}^2 \over r^2 ( r +  \tilde{\rho})^2} ,\nonumber\\
A_1 &=  { \tilde{\rho} \over 2 r^2 ( r + \tilde{\rho})}  \Bigl(  ( x^3 T^2 - x^2 T^3 ) +  e^{-2\phi_0}  \Theta_1 (x^1 T^1  +  x^2  T^2  +   x^3 T^3 )   \Bigr) ,  \nonumber\\
A_2 &=  { \tilde{\rho} \over 2 r^2 ( r + \tilde{\rho})}   \Bigl(  (  x^1T^3 -x^3T^1 ) +  e^{-2\phi_0}  \Theta_2 (x^1 T^1  +  x^2  T^2  +   x^3 T^3 )   \Bigr),  \nonumber\\
A_3&=   { \tilde{\rho} \over 2 r^2 ( r + \tilde{\rho}) }  \Bigl(  (x^2T^1 - x^1T^2)  +  e^{-2\phi_0}  \Theta_3 (x^1 T^1  +  x^2  T^2  +   x^3 T^3 ) \Bigr),  \nonumber\\
A_4&=  e^{-2\phi_0}  { \rho^2 \over 2 r^2 (\rho^2 + r) }   (x^1 T^1  +
 x^2  T^2  +   x^3 T^3 ),
\quad 
A_i = 0.
\end{align}
The solution seems complicated but one finds that the Yang-Mills gauge field satisfies the
anti-monopole equation \eqref{eq:anti-monopole}.
Notably, the function $\mathcal{H}$ becomes negative at a finite value
of $r$. 
The solution is ill-defined near the center of the brane. 
This property is similar to the symmetric and neutral NS5-brane
solutions with $n<0$ whose physical interpretation is unclear \cite{Gauntlett:1992nn}.

\subsection{Heterotic $5^2_2$-branes}
Now we are in a position where the second T-duality transformation is performed on the
KK5-brane solutions.
We introduce another isometry along the transverse direction $x^3$ to
the KK5-brane world-volumes.
By performing the second T-duality transformation along the $x^3$-direction, we obtain the exotic
$5^2_2$-brane solutions associated with the symmetric, the neutral and
the gauge solutions.

\paragraph{Symmetric $5^2_2$-brane}
The harmonic function $H(r)$ in $\mathbb{R}^2 \times T^2$ is 
obtained by the periodic array of the KK5-brane. This is given by 
\begin{align}
H (r) =& \ e^{2\phi_0} + \frac{1}{2} \sum_s \frac{n \alpha' R_3^{-1}}
{
\sqrt{r^2 + (x^3 - 2 \pi R_3 s)^2}
}, 
\notag \\
r^2 =& \ (x^1)^2 + (x^2)^2.
\label{eq:harmonic_sum2}
\end{align}
Taking the compactification radius small $R_3 \to 0$, 
the sum in \eqref{eq:harmonic_sum2} is approximated by the integral
over $s$ with the cutoff scale $\Lambda$. The result is
\cite{deBoer:2010ud}
\begin{align}
H (r) = h_0 - \frac{\sigma}{2} \log \frac{r}{\mu},
\label{eq:harmonic_func}
\end{align}
where $\sigma = \frac{R_3 R_4}{2 \pi \alpha'}$ and $\mu$ is a constant
which specifies the region where the solution is valid \cite{Kikuchi:2012za}.
The constant $h_0$ diverges in the large cutoff scale limit
$\Lambda \to \infty$. 
Since the harmonic function only depends on $x^1,x^2$, the NS-NS $B$-field
does so. 
The only non-zero component of the $B$-field $B_{34} = \omega$  
is determined by the following relations: 
\begin{align}
\partial_1 H = \partial_2 \omega, \qquad \partial_2 H = - \partial_1 \omega,
\end{align}
The explicit form of $\omega$ is found to be
\begin{align}
\omega = - \sigma \tan^{-1} 
\left(
\frac{x^2}{x^1}
\right).
\end{align}
If we employ the coordinate system $x^1 = r \cos \theta, x^2 = r \sin
\theta$ where $\theta$ is the angular coordinate in the $x^1x^2$-plane,
then $\omega$ is proportional to $\theta$.
This result leads to the fact that 
the space-time metric, the dilaton and the NS-NS $B$-field in the
solution are not single-valued anymore. 
The logarithmic behaviour of the harmonic function is characteristic to
co-dimension two objects.
We call the solution \eqref{eq:KKM_metric} on which the harmonic
function is replaced by \eqref{eq:harmonic_func} the smeared symmetric KK5-brane solution.
Indeed, the heterotic vortex and the domain wall are discussed in
\cite{Duff:1993ij,Onemli:2000kb} where the harmonic function behaves as logarithmic
and linear functions. 

Since the heterotic KK5-brane solution satisfies the 
standard embedding condition, this again simplify the heterotic Buscher
rule \eqref{eq:Buscher}. 
By performing the second T-duality transformation along the $x^3$-direction, we
obtain the following symmetric $5^2_2$-brane solution:
\begin{align}
ds^2 =& \ \eta_{ij} dx^i dx^j + H 
\left[
(dx^1)^2 + (dx^2)^2
\right] + \frac{H}{K} 
\left[
(dx^3)^2 + (dx^4)^2
\right], \notag \\
\phi =& \ \frac{1}{2} \log \frac{H}{K}, \quad 
B_{34} = - \frac{\omega}{K}, \quad 
K = H^2 + \omega^2.
\label{eq:symmetric_522}
\end{align}
The metric, the dilaton and the NS-NS $B$-field in
\eqref{eq:symmetric_522} are the same with the ones in type II theory.
However, in heterotic theory, the Yang-Mills gauge field remains non-trivial.
The gauge field $A_M {}^I$ is calculated as 
\begin{align}
A_1&=  { 1 \over 2H }  ( \partial_2 H  )T^3,   ~~~~~
A_2 = - { 1 \over 2H }  ( \partial_1 H  )T^3,   \nonumber\\
A_3&=   {1\over 2 HK} \Bigl(  (H \partial_2 H + \omega \partial_1 H )  T^1 -  (H \partial_1 H - \omega \partial_2 H )  T^2 \Bigr),   \nonumber\\
A_4 &=  - {  1 \over2 HK  }  \Bigl( (H \partial_1 H - \omega \partial_2
 H)T^1 + ( H \partial_2 H + \omega \partial_1 H )T^2   \Bigr),
\nonumber \\
A_i & = 0.
\label{eq:symmetric_522_gauge}
\end{align}
We find that \eqref{eq:symmetric_522_gauge} satisfies the 
vortex-like equation in two dimensions:
\begin{align}
& D_1 \varphi - i D_2 \varphi = 0, \qquad 
[\varphi, \bar{\varphi}] = B_3, 
\label{eq:vortex}
\end{align}
where $\varphi = \frac{1}{\sqrt{2}} (A_3 + i A_4)$ is a complexified
adjoint scalar field. The vortex-like equation \eqref{eq:vortex} is
obtained by dimensionally reducing the self-duality condition
\eqref{eq:self-dual} to two dimensions. 
We note that since the Yang-Mills gauge field contains the explicit angular
coordinate in $\omega$, $A_m$ is not a single-valued
function also in the symmetric $5^2_2$-brane solution.

The modified spin connection $\omega_{+}$ associated with the solution 
\eqref{eq:symmetric_522} 
is calculated as 
\begin{align}
& \omega_{+1} = \frac{1}{2 H} (\partial_2 H) T^3 
+ K^{-1}
\left(
H \partial_2 H + \omega \partial_1 H
\right) N^{34}, 
\quad
\omega_{+2} = - \frac{1}{2H}  (\partial_1 H) T^3 
- K^{-1} 
\left(
H \partial_1 H - \omega \partial_2 H
\right) N^{34}, 
\notag \\
& 
\omega_{+3} = - \frac{1}{2H}  K^{- \frac{3}{2}} 
\left(
\omega^2 \partial_2 H - H^2 \partial_2 H - 2 \omega H \partial_1 H
\right) T^1 + 
\frac{1}{2H}  K^{- \frac{3}{2}} 
\left(
\omega^2 \partial_1 H - H^2 \partial_1 H + 2 \omega H \partial_2 H
\right) T^2, 
\notag \\
&
\omega_{+4} = \ \frac{1}{2H}  K^{- \frac{3}{2}} 
\left(
\omega^2 \partial_1 H - H^2 \partial_1 H + 2 \omega H \partial_2 H
\right) T^1 
+ \frac{1}{2H}  K^{- \frac{3}{2}} 
\left(
\omega^2 \partial_2 H - H^2 \partial_2 H - 2 \omega H \partial_1 H
\right) T^2.
\end{align}
Here $(N^{ab})^{AB} = \delta^{aA} \delta^{bB} - \delta^{bA} \delta^{aB}$
is the generator of the $SO(4)$ Lorentz group. More explicitly, they are given by 
\begin{align}
& 
(N^{12}) = 
\left[
\begin{array}{cccc}
0 & 1 & 0 & 0 \\
-1 & 0 & 0 & 0 \\
0 & 0 & 0 & 0 \\
0 & 0 & 0 & 0
\end{array}
\right],
\quad 
(N^{13}) = 
\left[
\begin{array}{cccc}
0 & 0 & 1 & 0 \\
0 & 0 & 0 & 0 \\
-1 & 0 & 0 & 0 \\
0 & 0 & 0 & 0
\end{array}
\right],
\quad 
(N^{14}) =
\left[
\begin{array}{cccc}
0 & 0 & 0 & 1 \\
0 & 0 & 0 & 0 \\
0 & 0 & 0 & 0 \\
-1 & 0 & 0 & 0
\end{array}
\right],
\notag \\
& 
(N^{23}) = 
\left[
\begin{array}{cccc}
0 & 0 & 0 & 0 \\
0 & 0 & 1 & 0 \\
0 & 0 & 0 & 0 \\
-1 & 0 & 0 & 0
\end{array}
\right],
\quad 
(N^{24}) = 
\left[
\begin{array}{cccc}
0 & 0 & 0 & 0 \\
0 & 0 & 0 & 1 \\
0 & 0 & 0 & 0 \\
0 & -1 & 0 & 0
\end{array}
\right],
\quad 
(N^{34}) = 
\left[
\begin{array}{cccc}
0 & 0 & 0 & 0 \\
0 & 0 & 0 & 0 \\
0 & 0 & 0 & 1 \\
0 & 0 & -1 & 0
\end{array}
\right].
\end{align}
At first sight, the standard embedding condition
\eqref{eq:standard_embedding} does not hold for the symmetric $5^2_2$-brane
solution. However we find that the condition
\eqref{eq:standard_embedding} is satisfied up to the gauge
transformation. To see this, we calculate the gauge (and the $SO(4)$)
invariant quantity $\text{Tr} F_{MN} F^{MN}$ and $R_{MNAB} (\omega_+)
R^{MNBA} (\omega_+)$ and find 
\begin{align}
\text{Tr} F_{MN} F^{MN} = R_{MNAB} (\omega_+) R^{MNBA} (\omega_+).
\end{align}
This result indicates that $A_m$ and $\omega_+$ are identified up to a
gauge transformation. 

When the standard embedding 
condition is satisfied, all the $\alpha'$-corrections 
are expected to be canceled in the T-duality transformations \cite{Bergshoeff:1994dg}. 
We therefore conclude that the heterotic Buscher rule \eqref{eq:Buscher} is exact 
for solutions 
of the symmetric type. 
The family of solutions related by the T-duality transformations seem not
to suffer from the $\alpha'$-corrections. 
We note that the multi-valuedness which appears in the gauge invariant
quantity $\text{Tr} F_{MN} F^{MN}$ in the action is canceled by the standard embedding condition.
This is similar to the situation discussed in the multi-monopole
solutions \cite{Khuri:1992hk} where divergences in $\mathrm{Tr} F_{MN} F^{MN}$ are canceled in the term 
$R_{MNAB} R^{MNBA}$. 

\paragraph{Neutral $5^2_2$-brane}
For the neutral KK5-brane, we again find that the heterotic Buscher rule is
simplified due to the same reason discussed in the previous subsection.
After some calculations, the metric, the dilaton and the NS-NS $B$-field
are given by \eqref{eq:symmetric_522}.
This is a conceivable result since the neutral solution does not
involve gauge field anymore and it is a solution in type II theory.
We stress that although the symmetric solution \eqref{eq:symmetric_522} is an exact solution, 
the neutral solution is a perturbative solution in the $\alpha'$ expansion.

\begin{figure}[tb]
\centering
\includegraphics[width=14cm]{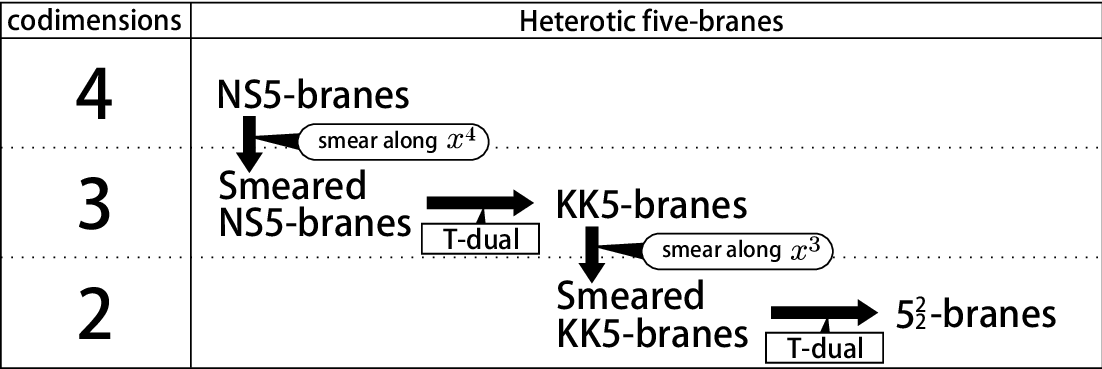}
\caption{
The smearing and the T-duality relations for the heterotic
 five-branes of various co-dimensions.
}
\label{fig:summary}
\end{figure}

\paragraph{Gauge $5^2_2$-brane} 
We introduce another isometry along the same way of obtaining the
Harrington-Shepard calorons for the smeared gauge NS5-brane.
The calculation is the same with the harmonic function
\eqref{eq:harmonic_func} in $\mathbb{R}^2 \times T^2$.
One obtains the function $f = \tilde{h}_0 - \frac{\tilde{\sigma}}{2}
\log \left( \frac{r}{\mu} \right) + \mathcal{O} (r/\Lambda)$ in the 't Hooft ansatz
\eqref{eq:tHooft}.
Here $r^2 = (x^1)^2 + (x^2)^2$ and 
$\tilde{h}_0 = 1 + \frac{\tilde{\rho}}{\pi R_3} \log [4 \pi R_3
\Lambda/\mu]$, $\tilde{\sigma} = \frac{2 \tilde{\rho}}{\pi R_3}$ are 
constants.
For the smeared gauge KK5-brane of co-dimension two, only the one
component of the $B$-field is non-trivial.
As discussed before, this is just a constant and we choose this $B_{34}
= \Theta$. The explicit form of the smeared gauge KK5-brane solution is
found in appendix.
Now we perform the second T-duality transformation of the gauge solution.
For the gauge KK5-brane, again the $\omega_{+}^2$ term in $G'_{MN}$ in
the heterotic Buscher rule is negligible as it is a higher order in
$\alpha'$.
However, the Yang-Mills gauge field remains non-trivial and contributes to the
Buscher rule. 
After calculations, we find the following gauge $5^2_2$-brane solution:
\begin{align}
ds^2 &=  \eta_{ij} dx^i dx^j + {\cal I }[( dx^1)^2 + ( dx^2)^2]  + { {\cal I}  \over e^{4\phi_0 }+\Theta^2} [( dx^3)^2 + ( dx^4)^2],   \nonumber\\
B_{34} &=- { \Theta \over e^{4\phi_0} + \Theta^2 },~~~~~
\phi_= {1\over2} \log ({ {\cal I} \over e^{4\phi_0} + \Theta^2 } ),~~~~~
{\cal I}=e^{2\phi_0}  -   { \alpha^{\prime} \tilde{\sigma}^2  \over 2 r^2 \bigl( \tilde{h}_0 -{ \tilde{\sigma} \over 2 } \log (r/\mu) \bigr)^2 }, \nonumber\\
A_1 &=  { - \tilde{\sigma} x^2  \over 4 r^2 \bigl( \tilde{h}_0 -{ \tilde{\sigma} \over 2 } \log (r/\mu) \bigr) }   T^3,~~~~~
A_2 =   {\tilde{\sigma} x^1  \over 4 r^2 \bigl( \tilde{h}_0 -{ \tilde{\sigma} \over 2 } \log (r/\mu) \bigr) } T^3, \nonumber\\
A_3 &=  { \tilde{\sigma}  \over4 r^2 \bigl( \tilde{h}_0 -{ \tilde{\sigma} \over 2 }\log (r/\mu) \bigr) ( e^{4\phi_0}+\Theta^2)}  \Bigl(   e^{2\phi_0} ( x^1 T^2- x^2 T^1) + \Theta ( x^1T^1 + x^2 T^2 )  \Bigr), \nonumber\\
A_4 &=  { \tilde{\sigma}  \over4 r^2 \bigl( \tilde{h}_0 -{
 \tilde{\sigma} \over 2 } \log (r/\mu) \bigr) ( e^{4\phi_0}+\Theta^2)}
 \Bigl(   e^{2\phi_0} ( x^1 T^1 + x^2 T^2)  +  \Theta ( x^1 T^2 - x^2
 T^1 )  \Bigr),
\nonumber \\
A_i  & =0.
\label{eq:gaugeKK5}
\end{align}
The gauge field satisfies the vortex-like equation
\eqref{eq:vortex} which is a reminiscent of the self-duality equation \eqref{eq:self-dual}.
In the symmetric and the neutral  
$5^2_2$-brane solutions, there was the function $\omega$
which contains explicit angular coordinate on the $x^1x^2$-plane.
In the gauge $5^2_2$-brane solution, the $B$-field is governed by a parameter $\Theta$
instead of $\omega$. 
However, since $\Theta$ is a constant parameter, the gauge $5^2_2$-brane 
solution is completely determined by single-valued functions and it
is a geometric solution. 
The function $\mathcal{ I }$ becomes negative at a finite value of $r$. 
This is the same situation in the case of the gauge KK5-brane and,
unfortunately, its physical meaning is still obscure. 
We will make a comment on this property in section 5. 
A summary of the smearing and the T-duality relations for the heterotic
five-branes is found in Figure \ref{fig:summary} .

\section{Monodromy and T-fold}
In this section, we study the monodromy of the heterotic $5^2_2$-brane
solutions.
The T-duality symmetry in heterotic string theories compactified on
$T^d$ with Wilson lines 
have been studied in detail \cite{Narain:1985jj, Narain:1986am}.
The Wilson line fields break the $SO(32)$ or $E_8 \times E_8$ gauge group
down to a Cartan subgroup of the gauge group. 
The off-diagonal parts of the Yang-Mills gauge field are Higgsed and 
becomes massive. 
In the lower-dimensions, there are $U(1)$ gauge fields which originate
from the Kaluza-Klein reduction of the metric, the NS-NS $B$-field and the
Yang-Mills gauge field that correspond to the Cartan subgroup. 
In this case, the T-duality group has been determined to be 
$O(d,d+\text{dim}G, \mathbb{R})$ where $\text{dim}G$ is the number of the $U(1)$
sector associated with the Cartan subgroup \cite{Maharana:1992my}.

On the other hand, when the Wilson line fields are absent, the non-Abelian gauge group is
not broken and the T-duality group $O(d,d+\text{dim}G,
\mathbb{R})$ reduces to $O(d,d,\mathbb{R})$ \cite{Hohm:2014sxa}.
Their study relies on the S-matrix analysis of strings and the result is
true in all orders in $\alpha'$.
In order to clarify the $O(d,d)$ covariance of field configurations, 
it is convenient to consider the generalized metric $\mathcal{H}$.
The generalized metric $\mathcal{H}$ in type II supergravity 
is a $2d \times 2d$ matrix and defined through the metric and the $B$-field.
Although, the T-duality group $O(d,d)$ itself does not change for all orders in $\alpha'$, 
the generalized metric receives $\alpha'$-corrections.
This is analogous to the $\alpha'$-corrected Buscher rule of the T-duality transformation.
Remarkably, in addition to the metric and the $B$-field, the Yang-Mills gauge field plays an
important role in heterotic theories.
The generalized metric in heterotic theories is determined by utilizing
the heterotic supergravity action \eqref{eq:heterotic_sugra} compactified on $T^d$
\cite{Hohm:2014sxa}. 
This is given by 
\begin{align}
\mathcal{H} = 
\left[
\begin{array}{cc}
G^{-1} & - G^{-1} B \\
B G^{-1} & G - B G^{-1} B
\end{array}
\right],
\label{eq:generalized_metric}
\end{align}
where $G$ is defined by 
\begin{align}
G_{\mu \nu} = g_{\mu \nu} + 2 \alpha' 
\bigl[
 \mbox{Tr} ( \omega_{+\mu}  \omega_{+\nu } )-  \mbox{Tr} ( A_{\mu}  A_{\nu} ) 
\bigr] ,
\label{eq:generalized_G}
\end{align}
Here $\mu, \nu$ are the isometry directions. 
The generalized metric \eqref{eq:generalized_metric} takes the same form
in type II supergravities but the second term in
\eqref{eq:generalized_G} is characteristic to heterotic theories.
The spin connection term in \eqref{eq:generalized_G} has been introduced
as it enters into the action \eqref{eq:heterotic_sugra} in the same way as the Yang-Mills gauge
field \cite{Hohm:2014sxa}.
In the following, we investigate the monodromy structures of the
heterotic $5^2_2$-branes by using the generalized metric \eqref{eq:generalized_G}.

\paragraph{Symmetric $5^2_2$-brane}
For the symmetric $5^2_2$-brane, since the standard embedding condition
is satisfied, we can choose a gauge where the second term in
\eqref{eq:generalized_G} is canceled.
We find that  
the generalized metric is the same with the one in type II theory.
For the solution \eqref{eq:symmetric_522}, this is given by
\begin{align}
\mathcal{H} (\theta) = 
\left[
\begin{array}{cccc}
H^{-1} K & 0 & 0 & H^{-1} \omega \\
0 & H^{-1} K & - H^{-1} \omega & 0 \\
  0   &  - H^{-1} \omega  & H K^{-1} + (HK)^{-1} \omega^2   & 0 \\
H^{-1} \omega & 0 & 0 & HK^{-1} + (HK)^{-1} \omega^2
\end{array}
\right].
\label{eq:symmetric_H}
\end{align}
When we go around the center of the $5^2_2$-brane and come back to the
original point, namely if the angular position changes as $\theta = 0 \to
2\pi$, then the generalized metric is evaluated as   
\begin{align}
\mathcal{H} (2\pi) = O^t \mathcal{H} (0) O,
\end{align}
where 
\begin{align}
O = \left[
\begin{array}{cc}
-i \tau_2 & 0 \\
2 \pi \sigma \mathbf{1}_2 & - i \tau_2
\end{array}
\right] \in O(2,2).
\end{align}
This implies that the monodromy is given by the $O(2,2)$ T-duality
transformation.
Therefore, although the symmetric $5^2_2$-brane solution is
non-geometric, it is a T-fold and a consistent solution to heterotic
string theories.

\paragraph{Neutral $5^2_2$-brane}
For the neutral $5^2_2$-brane, the bulk gauge field is trivial and we
can always choose the gauge where $A_M = 0$.
Again, the modified spin connection is
$\mathcal{O} (\alpha')$ and it does not contribute to
\eqref{eq:generalized_G} and to 
the generalized metric. 
Then the generalized metric is given by
\eqref{eq:symmetric_H} and its monodromy structure is the same with the
symmetric case. 
Therefore we find that the neutral $5^2_2$-brane is a T-fold 
at least at $\mathcal{O} (\alpha')$ in heterotic theories.

\paragraph{Gauge $5^2_2$-brane}
For the gauge $5^2_2$-brane, the situation is different.
The gauge field contributes to the generalized metric
through \eqref{eq:generalized_G} at $\mathcal{O} (\alpha')$.
However, since all the fields in the gauge solution do not depend on the angle $\theta$ in the
two-dimensional base space, they do not inherit multi-valuedness of the
geometry. Therefore the monodromy becomes trivial. 
This can be seen by evaluating the generalized metric for example in
$\Theta = 0$ gauge. In this gauge, we have
\begin{align}
\mathcal{H} = 
\left[
\begin{array}{cc}
G^{-1} & 0 \\
0 & G
\end{array}
\right],
\qquad
G = 
e^{-2\phi_0} 
\mathbf{1}_2.
\end{align}
This implies $ \mathcal{H} (2\pi) = \mathcal{H} (0) $.
Therefore we concludes that the gauge $5^2_2$-brane does not exhibit
non-geometric nature.

\section{Conclusion and discussions}
In this paper we studied the T-duality chains of five-branes in
heterotic supergravity.
A specific feature of heterotic supergravity is the Yang-Mills gauge sector
which appears in the linear order in the $\alpha'$-corrections.
There are also higher derivative corrections of the curvature square
term in the same order in $\alpha'$.
The three different half BPS five-brane solutions in this
$\alpha'$ order are known.
They are the symmetric, the neutral and the gauge NS5-brane solutions.
These are distinguished by the topological charge of instantons of
Yang-Mills gauge field and the charge associated with the modified $H$-flux.

We introduced the $U(1)$ 
isometry along a transverse direction to the NS5-brane
world-volume and explicitly performed the T-duality transformation of
these solutions.
Due to the $\alpha'$-corrections in heterotic supergravity, the Buscher
rule is modified by the corrections.
For the symmetric solution, where the standard embedding
condition is satisfied,
the $\alpha'$-corrections in the modified Buscher rule cancel out.
The resulting metric, the $B$-field and the dilaton are nothing but the ones for the KK5-brane
solution in type II theory.
We demonstrated 
that the Yang-Mills gauge field satisfies the standard embedding
condition again and it is given by the solution to the monopole equation
in three dimensions.
For the neutral solution, we find that the T-dualized solution is given by
the KK5-brane in type II theory at least at $\mathcal{O} (\alpha')$.
The solution is given by the purely geometric Taub-NUT metric. 
For the gauge solution, the geometry is ill-defined near the brane 
core after the smearing procedure. 
This property is carried over to the T-dualized solution.
For the gauge KK5-brane solution, the 
$B$-field is not excited which is
the same with the KK5-brane in type II theory.
However, the geometry is well-defined only at the asymptotic region.

We then introduce another $U(1)$ isometry to the KK5-brane solutions and 
perform the second T-duality transformation.
The resulting solutions are the exotic $5^2_2$-branes in heterotic
theory.
For the symmetric solution, the metric, $B$-field and the dilation are
given by that of the $5^2_2$-brane in type II theory.
The gauge field satisfies the vortex-like equation in two dimensions.
We found that the standard embedding condition is satisfied up to a
gauge transformation.
For the neutral solution, we found 
that the neutral $5^2_2$-brane is the
same with the one in type II theory.
They exhibit a non-geometric nature due to the multi-valuedness of the $B$-field.
For the gauge $5^2_2$-brane solution, we found that the fields do not show the
multi-valuedness and they remain geometric.

We next studied the monodromies of the three different $5^2_2$-branes. 
We calculated the generalized metric for these solutions.
We found that the symmetric and the neutral $5^2_2$-branes have 
a monodromy given by the $O(2,2)$ T-duality transformation. Therefore they are T-folds.
On the other hand, the gauge solution does not show the nature of a T-fold.

Following the general discussion in \cite{Bergshoeff:1994dg}, the
symmetric solution seems to be exact in terms of $\alpha'$.
On the other hand, for the neutral and the gauge $5^2_2$-brane solutions, 
they generically receive $\alpha'$-corrections.
The result is summarized in Table \ref{tb:522-branes}.
\begin{table}[tb]
\begin{center}
  \begin{tabular}{|l||l|l||l|} 
\hline
Type & Geometry & Valid order in $\alpha'$ & Notes \\ 
\hline \hline
Symmetric & T-fold & $\alpha'$-exact & standard embedding \\
Neutral & T-fold & $\mathcal{O} (\alpha')$ & Type II solution \\
Gauge & geometric & $\mathcal{O} (\alpha')$ & ill-defined near the center \\ 
\hline
  \end{tabular}
\end{center}
\caption{Properties of the heterotic $5^2_2$-branes.}
\label{tb:522-branes}
\end{table}

A few comments are in order about the solutions.
We introduced the $U(1)$ isometry to the gauge NS5-brane by the smearing
procedure of the instantons. 
The resulting gauge field is just the Harrington-Shepard calorons in
the small radius limit. The dilaton and the metric at $\mathcal{O}
(\alpha')$ are determined through the Bianchi identity $d \hat{H}^{(3)} =
\alpha' \mathrm{Tr} F \wedge F$ where the right-hand side is given by
the topological charge density for the smeared caloron. 
The resulting geometry is ill-defined near the center of the brane. 
We stress that there is another co-dimension three solution 
based on the BPS monopole of 't Hooft-Polyakov type instead of the smeared caloron \cite{Khuri:1992hk, Gauntlett:1992nn}. 
In the gauge NS5-brane solution of co-dimension three based on the BPS monopole type, 
the metric and the $B$-field behave well-defined near the core of
the five-brane. The gauge KK5-and $5^2_2$-branes of BPS monopole type would show better physical interpretation of T-dualized solutions.
A related property of the solutions is the logarithmic behaviour of the
$5^2_2$-branes of all types.
This is characteristic to the co-dimension two objects and found also in
the solution in type II theory \cite{deBoer:2010ud}.
Similar to the gauge solutions of the smeared caloron 
type, the $5^2_2$-branes discussed in section 3.2 seem to be ill-defined at
asymptotic region.
However this does not indicate any inconsistency of the solutions but
the general property of co-dimension two objects.
Analogous to the D7-brane in type IIB string theory, 
the exotic $5^2_2$-brane is not well-defined as the stand-alone object.
We need other co-existing branes in order to write down 
asymptotically flat globally well-defined solutions \footnote{An example
is the $SL(2,\mathbf{Z})$ multiplet of 7-branes in type IIB string
theory \cite{Greene:1989ya}.}.
Indeed, the scale $\mu$ in \eqref{eq:harmonic_func} specifies the
``cutoff'' point where the effect of the next duality branes is not
negligible \cite{Kikuchi:2012za}. 
We believe that globally well-defined $5^2_2$-brane solutions exist even
in heterotic theories.

We note that the most tractable way to study the non-geometric nature of
string theory solutions is the double field theory construction of
supergravity \cite{Berman:2014jsa, Bakhmatov:2016kfn}. 
There are several studies about double field theory formulation of 
heterotic supergravity \cite{Hohm:2011ex} and the inclusion of
$\alpha'$-corrections \cite{DFT}.
Although the spin connection term in the generalized metric
\eqref{eq:generalized_G} is a conjectural one, the supersymmetry
transformation law of the heterotic supergravity and 
the double field theory analysis strongly suggest that this is true \cite{Hohm:2014sxa}. 
Comprehensive studies are presented in the generalized geometry with
$\alpha'$-corrections \cite{Coimbra:2014qaa}.
The heterotic $5^2_2$-brane is expected to be a source of the non-geometric flux or
mixed-symmetric tensor and they are in the T-duality multiplets in
lower-dimensions \cite{Bergshoeff:2012jb}.
It is also interesting to study the world-volume effective action for the
non-geometric branes \cite{Chatzistavrakidis:2013jqa, Kimura:2014upa} in heterotic theory. 
We will come back to these issues in future studies.

\subsection*{Acknowledgments}
The authors would like to thank T.~Kimura and S.~Mizoguchi for useful discussions and comments.
The work of S.~S. is supported in part by Kitasato University Research Grant for Young
Researchers. The work of M.~Y. is supported by NUS Tier 1 FRC Grant R-144-000-316-112.

\begin{appendix}
\section{Smeared solutions for gauge type}
In this appendix, we introduce the explicit solutions of the smeared
gauge KK5-brane.
It is convenient first to introduce the defect gauge NS5-brane solution 
before we write down the smeared KK5-brane solution.
The defect gauge NS5-brane solution is obtained by the
smearing procedure along the $x^4$-direction to the $x^3$-smeared gauge
NS5-brane.
The resulting solution is a brane of co-dimension two.
By performing the T-duality transformation along the $x^3$-direction, we
obtain the $x^4$-smeared gauge KK5-brane.

\subsection{Defect gauge NS5-brane solution}
The defect gauge NS5-brane is the co-dimension two gauge NS5-brane and
it is obtained by smearing the two directions of the gauge
NS5-brane solution (\ref{eq:gauge_solution}) along the same way to
obtain the smeared gauge NS5-brane solution (\ref{SmearedD}).
The result is 
\begin{align}
ds^2 &=  \eta_{ij} dx^i dx^j + {\cal I } \delta_{ m n } dx^{m} dx^{n} ,    ~~~~~
{\cal I} = e^{2\phi_0}  - {  \alpha^{\prime}\tilde{\sigma}^2  \over 2 r^2 \bigl( \tilde{h}_0  -{ \tilde{\sigma}\over2 } \log (r/\mu) \bigr)^2 },  \nonumber\\
A_{m} &= - \bar{\sigma}_{mn} { \tilde{\sigma} x^{n} \over 4 r^2 \bigl( \tilde{h}_0  -{ \tilde{\sigma}\over2 } \log (r/\mu) \bigr) },   ~~~~
\phi= {1\over2} \log {\cal I}, ~~~
\hat{H}^{(3)}_{mnp} = -{1\over2} \varepsilon_{mnpq} \partial_{q} {\cal I}, \label{defectgauge}
\end{align}
where $r^2 = (x^1)^2+ (x^2)^2$. 
In the gauge solution, since the non-zero components of the 
modified $H$-flux $\hat{H}_{mnp}$ come from the Yang-Mills Chern-Simons
term, the $B$-field is taken to be a constant. 
For the defect NS5-brane solution, the relevant non-zero component of the $B$-field is $B_{34}=\Theta$. 
When we perform the heterotic T-duality transformation along the
$x^4$-direction for the defect NS5-brane solution, we can obtain the
smeared gauge KK5-brane solution shown below.

\subsection{Smeared gauge KK5-brane solution }
The smeared gauge KK5-brane solution is obtained by smearing
$x^3$-direction in the gauge KK5-brane solution (\ref{eq:gaugeKK5}).
The explicit form is as follows:
\begin{align}
ds^2 &=  \eta_{ij} dx^i dx^j + {\cal I } \delta_{ m^{\prime} n^{\prime} } dx^{m^{\prime}} dx^{n^{\prime}} + {\cal I}e^{-4\phi_0} [ dx^4 + \Theta dx^3 ]^2,   \nonumber\\
B_{MN}&=0,~~~~~
\phi= {1\over2} \log ( e^{-2\phi_0} {\cal I} ),~~~~~
{\cal I}=e^{2\phi_0}  -    {  \alpha^{\prime}\tilde{\sigma}^2  \over 2 r^2 \bigl( \tilde{h}_0  -{ \tilde{\sigma}\over2 } \log (r/\mu) \bigr)^2 },  \nonumber\\
A_1 &=   { -\tilde{\sigma} x^2  \over 4 r^2 \bigl( \tilde{h}_0  -{ \tilde{\sigma}\over2 } \log (r/\mu) \bigr)^2 } T^3 ,~~~~~
A_2  =  { \tilde{\sigma} x^1  \over4 r^2 \bigl( \tilde{h}_0  -{ \tilde{\sigma}\over2 } \log (r/\mu) \bigr)^2 } T^3, \nonumber\\
A_3 &=  {\tilde{\sigma}  \over 4 r^2 \bigl( \tilde{h}_0  -{ \tilde{\sigma}\over2 } \log (r/\mu) \bigr)^2 } \Bigl(   (  x^2 T^1 - x^1 T^2  )  +    e^{-2\phi_0} \Theta (  x^1 T^1 + x^2 T^2  )   \Bigr), \nonumber\\
A_4 &= e^{-2\phi_0}{ \tilde{\sigma}  \over4 r^2 \bigl( \tilde{h}_0  -{ \tilde{\sigma}\over2 } \log (r/\mu) \bigr)^2 } (x^1 T^1 + x^2 T^2). \label{smearedGaugeKKM} 
\end{align}
As we mentioned above, the solution is obtained by taking the heterotic T-duality transformation along the $x^4$-direction on the defect gauge NS5-brane solution (\ref{defectgauge}). 
When we take the heterotic T-duality transformation with the $x^3$-direction instead of the $x^4$-direction on (\ref{defectgauge}), we find the other type of smeared gauge KK5-brane solution. 
The solution is different with the sign in front of $\Theta$ in
(\ref{smearedGaugeKKM}), but the physical meanings are the same for both
of the solutions. 
On the other hands, 
if we take the heterotic T-duality along the $x^3$-direction for the smeared gauge KK5-brane, we obtain the gauge $5^2_2$-brane solution as we see in (\ref{eq:gaugeKK5}).
\end{appendix}

\end{document}